\documentclass[aps,prl,showpacs,amsmath,showkeys,twocolumn,floatfix,amssymb, preprintnumbers, nofootinbib, superscriptaddress]{revtex4} 
\usepackage{epsfig,dcolumn}
\usepackage{graphicx}
\usepackage{comment} 
\usepackage{verbatim}
\usepackage[usenames]{color}
\usepackage{bm}
 \usepackage{ifpdf}

\begin{document}

\title{ Transmission   coefficient of interacting few-body system in one dimensional  space}

\author{Peng~Guo}
\email{pguo@csub.edu}

\affiliation{Department of Physics,  California State University, Bakersfield, CA 93311, USA}
\affiliation{Kavli Institute for Theoretical Physics, University of California, Santa Barbara, CA 93106, USA}

\author{Vladimir~Gasparian}

\affiliation{Department of Physics,  California State University, Bakersfield, CA 93311, USA}

\date{\today}

\begin{abstract} 
We present a novel way of defining transmission coefficient of one spatial dimensional few interacting electrons system.  The formalism is based on the probability interpretation of  unitarity  of physical   scattering $S$-matrix. The relation of our formalism to the  well-established method for describing the conducting properties of non-interacting systems,  Landauer-B\"uttiker formula,  is discussed.   The transport properties of interacting two-electron system is also discussed as a specific example of our formalism. 
   \end{abstract}

\pacs{ }

\maketitle

%%%%%%%%%
%  Introduction  %
%%%%%%%%%

{\it Introduction.}---Few  particles interaction in one dimensional space (1D)   has attracted  a lot of attentions in various fields of physics  both experimentally \cite{Winkler:2006nat,Folling:2007nat,Moritz:2005prl,Kouwenhoven:2001rpp,Auslaender:2005sci,Fallani:2007prl} and theoretically \cite{Kane:1992prb,Matveev:1993prl,Fisher:MesoTran,Basko:2006ann,Deshpande:2010nat,Guo:2016fgl,Guo:2017crd,Guo:2018xbv}. The increased    interest in 1D system is largely   motivated by   recent advance on experimental techniques.  These new experimental techniques not only make studies of low-dimensional interacting particles experimentally possible, but also provide  opportunities to challenge our understanding of few-body systems. For instance,   a peculiar     bound state of atom pairs in an optical lattice with repulsive interaction between atoms has been observed recently  \cite{Winkler:2006nat}. Similar optical lattice technique is also   used to create  ultracold strongly interacting atoms  \cite{Fallani:2007prl}.  Single electron transistor and small quantum dots  may be other exciting   examples as     experimental realizable few-body systems \cite{Kouwenhoven:2001rpp}.     In such systems, the quantized energy levels and the Coulomb interaction  are comparable, so that only a small number of electrons can be confined to the region with the size of   orders of the Fermi wavelength.  It is hence commonly assumed that the electrons interact with each other only when they are on the same quantum dot  with contact interactions. Therefore, a set of nearly isolated small quantum dots may be considered as an ideal system for the studies of  few-body system, such as transport properties \cite{Wohlman:2000epl}.

Quantum transport  effect may play the crucial role in nanoscale semiconductor devices. The  transport properties of non-interacting electrons are usually described by Landauer-B\"uttiker formula \cite{Landauer:1970philmag,Buttiker:1985prb}, which establishes the relations between  the conductance $G$  of non-interacting electrons in a quasi-1D wire and  transmission  coefficient:  $G=\frac{e^2}{h}\sum_{n} \mathcal{T}_n$, where $ \mathcal{T}_n$ stand for the transmission  probability coefficient for the incoming state in  $n$-th channel. However, there is a firm belief that  the behavior of electrons in 1D wire may be strongly affected by  few-body interactions,  and even a small perturbation can  change the scattering pattern of particles significantly, see, {\it e.g.} Refs. \cite{Deshpande:2010nat,Guo:2016fgl,Guo:2017crd,Guo:2018xbv}.  For   interacting systems,   electron-electron interaction may be incorporated within Landauer-B\"uttiker formula  by including self-energy of electrons in a perturbation theory approach    \cite{Wohlman:2001prb,Ness:2010prb}. This approach is more or less   Hartree-Fock type approximation by assuming that electron-electron interactions are elastic on average, so that single electron maintains the same momentum and all the inelastic effects are included in self-energy of electron as virtual processes. Clearly, the approach may work well for many-body systems at larger scale, however, for few-body system, there are the cases that inelastic process may become essential \cite{Deshpande:2010nat,Guo:2016fgl,Guo:2017crd,Guo:2018xbv}. In this letter, we aim to present a general formalism for   describing transport properties of interacting few-body system. The formalism is based on the physical  transition $S$-matrix and  unitarity relation of few-body system, hence, it is suitable not only for weakly interacting few-body system, such as electron-electron coulomb interaction,  but also for strongly coupled systems that cannot be easily handled by perturbation theory.

{\it   $S$-matrix and wave packet prescription.}---In order to describe transport properties of a quantum conductor, one of the key elements is to introduce properly defined transmission or reflection coefficients to reflect the probabilities for electrons tunneling though potential barriers. Normally, for the single electron in 1D, the task may be easily accomplished by study the scattering solutions of Schr\"odinger equation. The asymptotical wave function of single electron in 1D has  quite a simple form with linear superposition of  two components: forward- and backward-going plane waves. Although, it has been a well-known fact that  the plane wave  is not normalizable and doesn't represent a physically realizable state. In the case of 1D single electron scattering, it is still possible to interpret the coefficients of two plane waves as physical probability amplitudes.  For example, assuming  a     particle incident  from left that is described by a forward-going plane wave:  $e^{i p x} $ \mbox{$ (p>0)$},  asymptotic wave function in forward and backward directions are given respectively by,
\begin{align}
\Psi(x, p)  & \stackrel{x \rightarrow + \infty}{ \longrightarrow}  \left [1+  i t(p    , p)  \right ] e^{i p x},  \nonumber \\
   & \stackrel{x \rightarrow - \infty}{ \longrightarrow}   e^{i p x} +   i t( -p    , p)   e^{-i p x},   \label{1pwave}
\end{align}
where  $  t(  k    , p)  $ represents the  scattering amplitude of particle.   The transmission and reflection probabilities are thus associated to   the   coefficients of transmitted wave in forward direction and reflected wave in backward direction:  \mbox{$\mathcal{T} = |1+ i t(p,p)|^2$} and  \mbox{$ \mathcal{R} = | i t(-p,p) |^2$} respectively.   Probability conservation yields the relation:  \mbox{$\mathcal{T}  + \mathcal{R} =1$}. Unfortunately, for multiple particles, even in one spatial dimension, the physical interpretation of coefficients in front of asymptotic wave functions  become problematic. The reason is that  the momenta   among particles after scattering may be redistributed, and the coefficients that describe the momenta redistribution   is not normalizable  due to plane wave description of aymptotic states. Using two-electron scattering off atoms as a example, with an incident plane wave of two electrons:  $e^{i \mathbf{ p} \cdot \mathbf{ x}} $, where \mbox{$\mathbf{ p} =(p_1,p_2)$} and \mbox{$\mathbf{ x} = (x_1,x_2)$} are 2D vectors that represent electron's momenta and positions respectively,  the      asymptotic form of two electrons wave function in forward direction is 
\begin{align}
& \Psi(\mathbf{ x}, \mathbf{ p})    \stackrel{\theta_x \rightarrow \theta_p}{ \longrightarrow   }  \left [ 2 \pi \delta(\theta_x - \theta_p)     + 2 i t ( p \mathbf{ \hat{x}} , \mathbf{ p})  \right ] \frac{e^{i (p x -\frac{\pi}{4})}  }{\sqrt{2 \pi p x}}    ,  \label{2pwave}
\end{align}
where \mbox{$\theta_x = \tan^{-1} \frac{x_2}{x_1}$}, \mbox{$\theta_p = \tan^{-1} \frac{p_2}{p_1}$}, \mbox{$\hat{\mathbf{ x}}=\frac{\mathbf{ x}}{x}$},   \mbox{$x=\sqrt{x_1^2+x_2^2}$},  and \mbox{$p=\sqrt{p_1^2+p_2^2}$}.  $t (  \mathbf{ p'} , \mathbf{ p})$ again  represents the scattering amplitude of two electrons. 
The expression in Eq.(\ref{2pwave}) resembles the one electron case in forward direction, however, the divergent  $\delta(\theta_x - \theta_p) $ term  in coefficient of outgoing spherical wave prevents direct physical interpretation of coefficient  as probability amplitude. The  divergence of   $\delta(\theta_x - \theta_p) $ term is due to the fact that  the angular component of   plane wave  which describes the momenta distribution among two electrons   is not normalizable. Plane wave description   violates Heisenberg uncertainty principle and is not suitable to be used to represent physical  states whose positions and momenta are both  well determined.

On the other hand,  it has been known that in  single electron scattering case given in Eq.(\ref{1pwave}), the coefficients of transmitted and reflected waves can be identified as reduced $S$-matrix elements after removing energy conservation $\delta$-function constraint,
\begin{equation}
 S(k,p)  = \delta_{ \hat{k},  \hat{p}} + i t(k,p),
\end{equation}
where \mbox{$\hat{k}= \frac{k}{|p|}$}, \mbox{$\hat{p}= \frac{p}{|p|}$} and \mbox{$|k|=|p|$}.   So that $\mathcal{T}= | S(p,p)|^2$ and $\mathcal{R} = | S(-p,p)|^2$, and probability conservation relation, $\mathcal{T}+ \mathcal{R}=1$, is equivalent to unitarity relation of $S$-matrix: 
\begin{equation}
\sum_{k = \pm p} | S(k,p) |^2  =1. \label{oneunit}
\end{equation}
 This connection is due to the fact that the transport behaviors    are determined by scattering properties of electrons in a conductor. Similarly, in case of two electrons, the coefficient    of   outgoing spherical waves, see Eq.(\ref{2pwave}),  can also be identified as reduced $S$-matrix element,
\begin{equation}
S( \mathbf{ k}, \mathbf{ p})  = 2\pi  \delta (\theta_k -  \delta_p )   +  2 i t( \mathbf{ k}, \mathbf{ p}) ,
\end{equation}
which satisfies unitarity relation,
\begin{equation}
 \oint \frac{ d \theta_k }{2\pi}   S^* (  \mathbf{ k} , \mathbf{ p})  S( \mathbf{ k}, \mathbf{ p}') = 2\pi \delta (\theta_p - \theta_{p'}). \label{unitarity}
\end{equation}
As already mentioned previously, in multiple particles cases, because of normalization issue of plane wave description of asymptotic states, the unitarity relation of $S$-matrix in Eq.(\ref{unitarity})  clearly cannot be interpreted directly as probability conservation relation.  In fact, based on   quantum scattering theory \cite{Eden:Smatrix},   the unitarity relation of   $S$-matrix corresponds to the physical requirement of probability conservation, only when  $S$-matrix element  indeed   represents the physical   probability amplitude for the occurrence of   transition. The transmission and reflection coefficients then may be introduced based on the probability interpretation of unitarity relation.

Above mentioned difficulty can be  remedied by introducing wave packet prescription of physical asymptotic states. The wave packet prescription is only required for the incident asymptotic state, since the outgoing states will be averaged out for physical transition process. Moreover, because only reduced $S$-matrix elements after removing energy conservation  constraint are used for the transport properties, it is sufficient to introduce  the wave packet  of multiple particles that only describes the   momenta distribution among particles corresponding to a fixed total energy. Using again two electrons case as a example, the physical wave function with a incoming wave packet of two electrons may be defined by
\begin{equation}
\Phi(\mathbf{ x}, \mathbf{ p}_0) =  \oint \frac{d\theta_p}{2\pi}\Psi(\mathbf{ x}, \mathbf{ p}) \Theta(\theta_p, \theta_{p_0}), 
\end{equation}
where the function $\Theta  (\theta_p, \theta_{p_0})$ describes  the angular distribution of incoming wave packet peaked around a  fixed angle $\theta_{p_0}$, and is normalized according to relation,
  \begin{equation}
  \oint \frac{d\theta_p}{2\pi}  \left |  \Theta  (\theta_p, \theta_{p_0}) \right |^2  =1.
  \end{equation}
  Hence, the physical  $S$-matrix element  is related to plane wave based $S$-matrix element by,
  \begin{equation}
 \mathcal{S} (  \theta_k ,  \theta_{p_0})  =   \oint \frac{d\theta_p}{2\pi}    S(\mathbf{ k}, \mathbf{ p})  \Theta  (\theta_p, \theta_{p_0}) .  \label{physmat} 
\end{equation}
The physical $S$-matrix element in Eq.(\ref{physmat}) and   the unitarity relation in Eq.(\ref{unitarity}) together yields a well-defined  probability conservation relation,
\begin{equation}
 \oint \frac{ d \theta_k }{2\pi}    |  \mathcal{S} ( \theta_k ,  \theta_{p_0})  |^2=1. \label{phyunit}
\end{equation}
Therefore, $ |  \mathcal{S} ( \theta_k ,  \theta_{p_0})  |^2$   now can be used to represent the probability of transition from initial momenta configurations peak at $\theta_{p_0} \in [0,\frac{\pi}{2}]$ into final configuration  $\theta_k$.  The transmission coefficient  may be introduced by considering all configurations scattered into forward direction,
 \begin{equation}
\mathcal{T}  = \int_0^{\frac{\pi}{2}} \frac{ d \theta_k }{2\pi}    |  \mathcal{S} ( \theta_k ,  \theta_{p_0})  |^2  , \label{phytranscoef}
\end{equation}
and based on the probability conservation, the reflection coefficient thus is simply given by $\mathcal{R} =1-\mathcal{T} $.

The    arguments  and wave packet prescription can  be   generalized  to multiple electrons as well. The plane wave based unitarity relation  for $D$ electrons has the similar form as in Eq.(\ref{unitarity}),
\begin{equation}
\oint \frac{d \hat{\mathbf{ k}}}{(2\pi)^{D-1}} S^{\dag}(\mathbf{ k}, \mathbf{ p}) S (\mathbf{ k}, \mathbf{ p}') = (2\pi)^{D-1} \delta^{D-1}  (\hat{\mathbf{ p}}-   \mathbf{ \hat{p}}') ,
\end{equation}
where $S (\mathbf{ k}, \mathbf{ p})$ stands for the plane wave based reduced $D$-electron scattering $S$-matrix. The  incoming and outgoing electron's momenta are represented by   $D$-dimensional space vectors,  $\mathbf{ p}=(p_1,  \cdots , p_D)$ and $\mathbf{ k}=(k_1,  \cdots , k_D)$,       and they  are constrained by energy conservation: $k^2=p^2$. Hence,  unit vectors in $D-1$ dimensional space,      $\mathbf{ \hat{p}} =\frac{\mathbf{ \hat{p}}}{p}$ and $\mathbf{ \hat{k}} =\frac{\mathbf{ \hat{k}}}{p}$, may be used to define incoming and outgoing electron's internal momenta distribution, which  resembles the angular distribution in $D-1$ dimensional space mathematically. By introducing  a normalized angular function $\Theta ( \mathbf{ \hat{p}} , \mathbf{ \hat{p}}_0 )$ to describe the wave packet of incoming particles, 
the transmission coefficient of $D$-electron may  be defined again based on $D$-electron probability conservation relation,
  \begin{equation}
\oint \frac{d \hat{\mathbf{ k}}}{(2\pi)^{D-1}}  | \mathcal{S} (\mathbf{ \hat{k}}, \mathbf{ \hat{p}}_0)|^2 = 1 ,
\end{equation}
where  physical $S$-matrix element of $D$-electrons is given by
\begin{equation}
 \mathcal{S} (\mathbf{ \hat{k}}, \mathbf{\hat{ p}}_0) =\oint \frac{d \hat{\mathbf{ k}}}{(2\pi)^{D-1}}  S (\mathbf{ k}, \mathbf{ p})\Theta ( \mathbf{ \hat{p}} , \mathbf{ \hat{p}}_0 ) . 
\end{equation}

{\it  Two electrons interaction   in a crystal.}---In this section,  as a specific example of our formalism, we use a simple solvable  model  to demonstrate some interesting  features and  transport properties of two interacting electrons in crystal.   The dynamics of our model is  given by Hamiltonian,
\begin{equation}
\hat{H}  = \hat{T} + V(x_1) + V(x_2) +U(\mathbf{ x} ),
\end{equation}
where $\hat{T} = - \frac{1}{2m} \sum_{i=1,2} \frac{d^2}{d x^2_i}$ is kinetic energy of two electrons. Two types of interactions are considered in this model: (1) $V(x_i)$ describes the pair-wise interaction between $i$-th electron and  atoms in crystal, and  (2) $U(\mathbf{ x} )$ represent the three-body interaction involving both electrons and atoms in crystal.  $U$-type three-body potential may be approximated by   contact interactions,
\begin{equation}
U(\mathbf{ x}) = \sum_{\alpha=0}^{N-1} U_0 \delta (x_1 - a_\alpha) \delta(x_2 - a_\alpha),
\end{equation}
where $a_\alpha$ are the locations of atoms and $N$ denotes the total number of atoms, hence $U$-type potential only contribute when both electrons meet at same atom.  The spin effect has been neglected in this model, and   solutions for spin triplet and singlet may be achieved by symmetrization of wave function. The scattering solution of Schor\"odinger equation for two electrons system   may be obtained by considering  Lippmann-Schwinger equation, see  {\it e.g.} \cite{Guo:2017ism}.  The $S$-matrix may be introduced  by studying the asymptotic wave function of two electrons system. In our case,  asymptotic wave function is given by combination of both  plane waves   and spherical wave   \cite{Guo:2017ism},
\begin{align}
\Psi(\mathbf{ x}, \mathbf{ p}) & \rightarrow    \left [e^{i p_1 x_1}  + i t(p_1 \hat{x}_1 ,p_1) e^{i p_1 | x_1 |} \right ] \nonumber \\
& \quad  \times   \left [e^{i p_2 x_2}  + i t(p_2 \hat{x}_2  ,p_2) e^{i p_2 | x_2 |} \right ] \nonumber \\
& + 2 i T_U ( p \mathbf{ \hat{x}} , \mathbf{ p})    \frac{e^{i (p x -\frac{\pi}{4})}  }{\sqrt{2 \pi p x}}  , \label{asympwave}
\end{align}
where $\hat{x}_i = \frac{x_i}{|x_i|}$, and  $t(p_i \hat{x}_i, p_i)$ denotes for the scattering amplitude of single electron.  The three-body  scattering amplitude, $T_U$, is given by 
\begin{equation}
T_U( \mathbf{ k}, \mathbf{ p}) = - \frac{m}{2}   \sum_{\alpha,\beta=0}^{N-1}  \phi^*  (\mathbf{ a}_{\alpha}, \mathbf{ k})    \left [ \mathcal{D}^{-1} \right ]_{\alpha, \beta}  \phi  (\mathbf{ a}_{\beta} , \mathbf{ p}) ,
\end{equation}
where    $\mathbf{ a}_\alpha = (a_\alpha, a_\alpha)$, and     $\mathcal{D}$  matrix is  defined as
\begin{equation}
\mathcal{D}_{\alpha,\beta} = \frac{1}{U_0} \delta_{\alpha,\beta} - G (\mathbf{ a}_\alpha, \mathbf{ a}_{\beta}).
\end{equation}
The   wave function $\phi (\mathbf{ x} , \mathbf{ p})$ and Green's function $G(\mathbf{ x}, \mathbf{ x}') $  satisfy equations,
\begin{equation}
\left [  \hat{T} + V(x_1) + V(x_2) \right ]  \phi  (\mathbf{ x} , \mathbf{ p}) =E \phi   (\mathbf{ x}, \mathbf{ p} ) ,
\end{equation}
and 
\begin{equation}
\left [ E- \hat{T} - V(x_1) - V(x_2) \right ] G (\mathbf{ x}, \mathbf{ x}') = \delta( \mathbf{ x} - \mathbf{ x}' ),
\end{equation}
 respectively.
 
The plane waves in Eq.(\ref{asympwave}) are the result of  pair-wise $V$-potentials between electrons and atoms in crystal. The coefficient of spherical wave, $T_U$, stands for the "true"  three-body $U$-type interaction when both electrons and atoms are involved in interaction. 
After removing energy conservation constraint, the reduced plane wave basis $S$-matrix are given by
\begin{equation}
S( \mathbf{ k}, \mathbf{ p})  =\sum_{\delta = \pm \theta_p}^{\pi \pm \theta_p  }  2\pi \delta (\theta_k -  \delta ) s_V (\mathbf{ k} , \mathbf{ p} )  +  2 i T_U( \mathbf{ k}, \mathbf{ p}) ,
\end{equation}
where $\theta_k =\tan^{-1} \frac{k_2}{k_1}$, $\theta_p =\tan^{-1} \frac{p_2}{p_1}$, and
\begin{equation}
 s_V (\mathbf{ k} , \mathbf{ p} ) = \left [ \delta_{ \hat{k}_1 ,  \hat{p}_1 } + it (k_1,p_1) \right ]  \left [ \delta_{ \hat{k}_2 , \hat{p}_2 }   +   it (k_2,p_2) \right ],
\end{equation}
stands for the reduced $S$-matrix   with only the presence of $V$-potentials, where $\hat{k}_i = \frac{k_i}{|p_i|}$ and  $\hat{p}_i = \frac{p_i}{|p_i|}$.     The reduced   plane wave basis $S$-matrix satisfies  unitarity relation given in Eq.(\ref{unitarity}). 
A simple choice of angular distribution function $\Theta  (\theta_p, \theta_{p_0})$ may be a gaussian function, 
\begin{equation}
\Theta  (\theta_p, \theta_{p_0}) =   \frac{(2\pi)^{\frac{1}{4}}   }{ \sqrt{ \tau } } e^{- \frac{(\theta - \theta_{p_0})^2}{4 \tau^2}},
\end{equation}
where $\tau$ is a small parameter to control the width of peak of wave packet.  Hence the physical $S$-matrix that describes the physical transition probability of two electrons  is given by
\begin{align}
 \mathcal{S} (  \theta_k ,  \theta_{p_0}) & =  \sum_{\theta_p = \pm \theta_k}^{\pi \pm \theta_k  }     s_V (\mathbf{ k} , \mathbf{ p} ) \Theta  (\theta_p, \theta_{p_0})   \nonumber \\
 &+  2 i   \oint \frac{d\theta_p}{2\pi}    T_U( \mathbf{ k}, \mathbf{ p})   \Theta  (\theta_p, \theta_{p_0}) ,
\end{align}
and the transmission coefficient   defined by Eq.(\ref{phytranscoef})     represents the probability of finding both two electrons scattered into forward direction.

\subsection{$U=0$ limit}
 In the case of $U=0$,  the $T_U  (\mathbf{ k} , \mathbf{ p} )  \rightarrow 0$,    the unitarity relation of physical $S$-matrix in Eq.(\ref{phyunit}) is thus reduced to a simple form,
 \begin{equation}
  \sum_{k_i=\pm p_i}  |    s_V (\mathbf{ k} , \mathbf{ p} )|^2 =1, \label{sVoptic}
\end{equation}
this is exactly what we expected for non-interacting two electrons.   Assuming \mbox{$(p_1 >0, p_2>0) $}, the transmission and reflection coefficients may be introduced by
\begin{align}
\mathcal{T}  & = \left | s_V( p_1, p_2 ;  p_1, p_2) \right |^2  , \nonumber \\
\mathcal{R}  & = \left | s_V( -p_1, p_2 ;  p_1, p_2) \right |^2 +  \left | s_V( p_1,- p_2 ;  p_1, p_2) \right |^2  \nonumber \\
&  + \left | s_V( -p_1,  -p_2 ;  p_1, p_2) \right |^2.
\end{align}
Hence, $\mathcal{T}  $ does indeed describe the probability of finding  both  electrons    in forward direction after scattering.
The complication in reflection coefficient of two electrons are due to the fact that  two electrons wave function now has four independent plane waves: \mbox{$e^{ \pm i p_1 x_1}  e^{ \pm i p_2 x_2} $}, in addition to both electrons in forward direction, which create     three  other scenarios: (i)  \mbox{$s_V( -p_1, p_2 ;  p_1, p_2)  $}   describe particle-2 moves forward and particle-1 is scattered backward; (ii) similarly,  \mbox{$s_V( p_1,-  p_2 ;  p_1, p_2) $}  is related to particle-1 moves forward and particle-2 is scattered backward;  (iii) and \mbox{$s_V( -p_1, -p_2 ;  p_1, p_2) $} is associate with both particles are scattered backward.

\subsection{$V=0$ limit}
At another extreme limit, as \mbox{$V \rightarrow 0$}, \mbox{$s_V (\mathbf{ k} , \mathbf{ p} ) \rightarrow \delta_{\hat{k}_1, \hat{p}_1}\delta_{\hat{k}_2, \hat{p}_2}$},  \mbox{$\phi (\mathbf{ x} , \mathbf{ p}) \rightarrow e^{i \mathbf{ p} \cdot \mathbf{ x}}$}, and 
\begin{equation}
G  (\mathbf{ x}, \mathbf{ x}') \rightarrow - \frac{  m }{2} i H_0^{(1)} ( p |  \mathbf{ x}- \mathbf{ x}'|).
\end{equation}
   The physical $S$-matrix now has a  form,
\begin{align}
 \mathcal{S} &(  \theta_k ,  \theta_{p_0})  = \Theta  (\theta_k, \theta_{p_0})   +  2 i \oint \frac{d\theta_p}{2\pi} T_U( \mathbf{ k}, \mathbf{ p})\Theta  (\theta_p, \theta_{p_0})  ,
\end{align}
where
  \begin{equation}
T_U( \mathbf{ k}, \mathbf{ p}) = - \frac{m}{2}    \sum_{\alpha,\beta=0}^{N-1}  e^{- i\mathbf{  k} \cdot \mathbf{ a}_\alpha }  \left [ \mathcal{D}^{-1} \right ]_{\alpha, \beta}  e^{ i\mathbf{  p} \cdot \mathbf{ a}_\beta }   ,
\end{equation}
and
\begin{equation}
\mathcal{D}_{\alpha,\beta} = \frac{1}{U_0} \delta_{\alpha,\beta}+ \frac{m}{2} i H_0^{(1)} ( p |  \mathbf{ a}_\alpha -  \mathbf{ a}_\beta|).
\end{equation}
The diagonal matrix elements $\mathcal{D}_{\alpha,\alpha} $ present another difficulty due to the ultraviolet divergence of Hankel function  at   origin,
\begin{equation}
H_0^{(1)} (p  r ) \stackrel{r \rightarrow 0}{\rightarrow}  1+ \frac{2 i}{\pi} \left (\gamma_E + \ln \frac{\ p   }{\Lambda} \right ) , \ \  \ \ \Lambda = \frac{2}{r},
\end{equation}
where $\Lambda$ is served as ultraviolet regulator.  Ultimately, the physical result should not depend on the choice or regulator and it will be set to \mbox{$\Lambda \rightarrow \infty$}.
The ultraviolet divergence may be dealt with standard renormalization procedure \cite{Cavalcanti:1998jx,Mitra:1998vr}. The ultraviolet divergence in Hankel function at origin may   be absorbed by bare coupling strength $U_0$,   a scale dependent running renormalized coupling strength is hence introduced by
\begin{equation}
\frac{1}{U_R(\mu)}  = \frac{1}{U_0} -  \frac{ m}{\pi} (\gamma_E + \ln \frac{ \mu    }{\Lambda})   ,
\end{equation}
where $\mu$ stands for the renormalization scale, and   $U_R(\mu)$ is the  physical coupling strength measured at scale $\mu$. The diagonal matrix element $\mathcal{D}$ is now given by
\begin{equation}
\mathcal{D}_{\alpha,\alpha} =  \frac{m}{2} i  + \frac{1}{U_R(\mu)}  - \frac{m}{\pi} \ln \frac{p }{ \mu }. \label{Dscaledep}
\end{equation}
The physical observable,  $\mathcal{D}$,  shouldn't depend on the renormalization scale $\mu$, 
\begin{equation}
\frac{d }{ d \mu } \mathcal{D}_{\alpha,\alpha}  =0.
\end{equation}
Hence it   yields a equation for running coupling strength,
\begin{equation} 
  \frac{d U_R (\mu)}{d  \ln \mu} = \frac{m U_R^2(\mu)}{\pi} ,
\end{equation}
and  the solution of running coupling strength is given by
\begin{equation}
\frac{1}{m U_R(\mu)}  = \frac{1}{m U_R^B  }    - \frac{1 }{\pi}   \ln \frac{\mu}{\mu_B} , \label{runUR}
\end{equation}
where the initial condition of physical observable \mbox{$U^B_R = U_R (\mu_B) $} is coupling strength measured at scale $\mu_B$. The scale dependence in $U_R(\mu)$ and  \mbox{$\frac{m}{\pi} \ln \frac{p }{ \mu }$}  in Eq.(\ref{Dscaledep}) cancel out, so ultimately, physical observable, $\mathcal{D}$, indeed doesn't  depend on the choice of renormalization scale $\mu$: 
\begin{equation}
\mathcal{D}_{\alpha,\alpha}   =   \frac{m}{2} i   +\frac{1}{U_R^B}  - \frac{m}{\pi} \ln \frac{p }{\mu_B } .  
\end{equation}

 For the weak  coupling   (\mbox{$U^B_R \ll m^{-1}$}),  the $\mathcal{D}$ matrix may be approximated by only   diagonal elements: \mbox{$\mathcal{D}_{\alpha,\alpha}  \sim  \delta_{\alpha , \beta}  \frac{1}{U_R^B}$}, hence,  
\begin{align}
T_U(  \mathbf{ k}, \mathbf{ p}) & \rightarrow   -  \frac{\frac{1}{2}}{ \frac{i}{2}  + \frac{1}{m U_R^B}  - \frac{1}{\pi} \ln \frac{p }{ \mu_B } } 
 \frac{e^{i\frac{ p L\Omega}{\sqrt 2 }}}{e^{i\frac{p L\Omega}{ \sqrt 2 N}}}
\frac{\sin{\frac{  p L \Omega}{\sqrt 2}}}{\sin{\frac{p L \Omega}{ \sqrt 2 N}}},
\end{align}
where $\Omega\equiv \cos(\theta_p-\frac{\pi}{4})-\cos(\theta_k-\frac{\pi}{4})$.  We  have also assumed that all atoms  are separated with even distance: \mbox{$a_{\alpha} = \frac{L}{N} \alpha$}, $\alpha=0,\cdots, N-1$, where $L$ stands for the length of crystal.  The transmission coefficient, in case of a shape peaked wave packet ($\tau \rightarrow 0$), is now given by
\begin{equation}
\mathcal{T} \rightarrow 1-\frac{{2\tau}}{{(2\pi)}^{\frac{3}{2}}}\frac{1}{ \frac{1}{4}  +  (\frac{1}{m U_R^B}  - \frac{1}{\pi} \ln \frac{p }{ \mu_B })^2 }\int^{2\pi}_{\frac{\pi}{2}}\frac{\sin^2{\frac{p L \Omega}{\sqrt 2}}}{\sin^2{\frac{p L \Omega}{ \sqrt 2 N}}}d\theta_k. \label{T}
\end{equation}
One of interesting feature in two interacting electrons case is that due to the $U$-type three-body interaction,  the integrand expression in Eq.(\ref{T}),  $\frac{\sin^2{\frac{  p L \Omega}{\sqrt 2}}}{\sin^2{\frac{p L \Omega}{ \sqrt 2 N}}}$,   shows the interference pattern and resembles  to intensity distribution from an ideal grating with $N$ slits in optics or  the resistance of one-dimensional chains in Kronig-Penny-like models (see, e.g. \cite{vg88}). In contrast, in the case of  the single electron   interacting with  $N$ numbers of contact interactions,  even at weak coupling limit,     the phase factors in forward scattering amplitude all cancel out. The transmission coefficient for single electron is independent of phase factors:  $ \mathcal{T}=1- N^2 \frac{m^2 U_0^2}{p^2}$, and shows no interference pattern.

{\it  Discussion and Summary.}---In order to see the resemblance of multiple channels Landauer-B\"uttiker formula and multiple particles $S$-matrix formalism, let's   consider  the case of single electron traveling in a quasi-1D wave guide along $z$-direction.  The potential barrier is placed at center of wave guide, and the motion of electron in transverse direction is confined in a narrow tube. Hence, the energy spectra  in transverse direction is discretized, the wave function is given by the product of a plane wave in $z$-direction, $e^{i p_n z}$, and bound state wave function in transverse direction, $ \Phi_{n} (x,y) $, where $n$ refers to the $n$-th energy state, $\epsilon_n$, in transverse direction, and $p_n = \sqrt{2m (E-\epsilon_n)}$. Assuming initial incident electron is in $n$-th eigenstate, thus,  the scattered wave function of electron  in forward direction is given by
 \begin{equation}
 \Psi_n(\mathbf{ x},E) \rightarrow \sum_{n'} S_{n,n'} \Phi_{n'} (x,y) e^{i p_{n'} z},
 \end{equation}
 where $ S_{n,n'}$ is scattering $S$-matrix element between $n$-th and $n'$-th  channels, and satisfies unitarity relation: $\sum_{n'} |S_{n, n'}|^2 =1$.  Since   transverse wave function, $ \Phi_{n} (x,y) $, is also well normalized according to
 \begin{equation}
 \int d x d y \Phi^*_{n'} (x,y) \Phi_{n} (x,y) = \delta_{n,n'},
 \end{equation}
 the coefficient of plane wave in $z$-direction, $S_{n,n'} \Phi_{n'} (x,y)$,   may still be used  to describe probability of physical transition process. Hence, the transmission coefficient in initial channel-$n$  may be defined as net result of coefficient square,
  \begin{equation}
\mathcal{T}_n  =  \int d x d y  |  \sum_{n'} S_{n,n'} \Phi_{n'} (x,y) |^2  = \sum_{n'} | S_{n,n'}   |^2   .
 \end{equation}
 
 In the case of two electrons, the situation is somehow similar,   the two electrons wave function in  forward direction is now described by  outgoing spherical waves, $\frac{e^{  i (px -\frac{\pi}{4})}}{\sqrt{2\pi px}}$, propagating in radial direction, and the angular dependent  physical $S$-matrix element,
  \begin{equation}
\Phi(\mathbf{ x}, \mathbf{ p}_0) \stackrel{\theta_x \rightarrow \theta_{p_0}}{ \longrightarrow }  \mathcal{S} (\theta_x, \theta_{p_0} )  \frac{e^{i(px - \frac{\pi}{4})}}{\sqrt{2\pi p x}}   . \label{wavepaket}
\end{equation}
 If each possible configuration of allowed momenta distribution among particles is labelled as a single channel, in multiple particles case,   there are infinite  channels.  The scattering of multiple particles may be treated as a continuously distributed multiple-channel problem. Physical $S$-matrix element square, $| \mathcal{S} (\theta_x, \theta_{p_0} )|^2$, hence describe the transition probability between channel-$\theta_x$ and channel-$\theta_{p_0} $.  The transmission coefficient in initial channel-$\theta_{p_0}$  is thus given by   net result of all   forward transitions,
 \begin{equation}
\mathcal{T}_{ \theta_{p_0}}  =   \int_0^{\frac{\pi}{2}} \frac{d\theta_x}{2\pi}   | \mathcal{S} (\theta_x, \theta_{p_0} ) |^2  .
\end{equation}

In summary, the  transport properties of few-electron system is normally complicated by some new features due to multiple particles interaction effect, such as interference and diffraction. The proper approach of introducing transmission and reflection coefficient of few-electron system is discussed in present work based on the probability interpretation of physical unitarity relation of  scattering $S$-matrix. The normalization  paradox of unitarity relation is remedied by the wave packet description of incident physical states.

 {\it Acknowledgement.}---We thank B.~Altshuler, M.~Ortuu{\~n}o and E.~Cuevas for their useful comments. 
 P.G. acknowledges partial support by the National Science Foundation under Grant No. NSF PHY-1748958.

\end{document}